# Record High Polarization at 2V and Imprint-free operation in Superlattice $HfO_2$-$ZrO_2$ by Proper Tuning of Ferro and Antiferroelectricity

Xinye Li, Sayani Majumdar*

Information Technology and Communication Sciences, Tampere University, 33720 Tampere, Finland

*Email: sayani.majumdar@tuni.fi

Neuromorphic computing, inspired by biological intelligence, offers a pathway to revolutionize artificial intelligence (AI) by unifying memory and processing in an energy-efficient, sustainable framework for data-intensive tasks. Ferroelectric (FE) materials have emerged as promising candidates for implementing artificial synapses, yet achieving low-voltage operation in CMOS back-end compatible devices remains a major challenge. In this work, we demonstrate that proper tuning of ferro and antiferroelectric phase in $HfO_2$–$ZrO_2$ (HZO) superlattice based capacitors can lead to imprint-free switching with record switchable polarization ($2P_r$) of 76 µC/cm² under an external field of only 2 MV/cm. The sizable remanent polarization of the superlattice HZO further enables linear potentiation and depression with an on/off ratio of 20 within a 3 MV/cm bias window. Under pulsed operation, the devices show robust endurance, either maintaining polarization with less than 10% degradation up to $10^8$ cycles or surviving beyond $10^9$ cycles with recoverable fatigue. By elucidating two distinct fatigue mechanisms, this work highlights strategies for optimizing FE devices to meet the stringent demands of neuromorphic training applications.

**Key Points**: low power, imprint-free, high polarization, analog control, recovery and endurance

# 1. Introduction

In the era of data-intensive computing, the importance of energy-efficient, fast and robust non-volatile memory (NVM) technology has become enormous.[1, 2] For the development of bio-inspired neuromorphic electronics, devices and systems aim to combine memory and processing functionalities within a single architecture. For doing artificial intelligence (AI) tasks in a more energy-efficient and sustainable way, new hardware with synaptic and neuronal functionalities needs to be developed that will support more high-performance electronic systems compared to conventional digital hardware.[3, 4, 5] In the large-scale neuromorphic systems, the implementations remain fundamentally digital and thus are not sufficiently energy efficient. For true advancement, analog neuromorphic hardware, integrated with mature complementary metal-oxide-semiconductor (CMOS) technology, is required. This integration demands materials and device architectures that meet back-end-of-line (BEOL) process compatibility, device stability, long operational lifetimes, and cost-effective large-scale manufacturability. Among different emerging memory technologies, ferroic material based magnetic random-access memory (MRAM) and ferroelectric (FE) memories form a very promising class due to their stable switching over a large temperature range, energy efficiency and CMOS BEOL integrability.[6] FE materials show particularly promising features for implementing analog synaptic computation due to controllable intermediate remanent polarization ($P_r$) states. This fine-tuned non-volatile polarization states enable both memory functions [7] and programmable synaptic. [3, 5, 8] Alongside these functionalities, FE materials also exhibit piezoelectricity and pyroelectricity, further expanding their potential applications in next-generation electronics combining sensing and computing.

Recently, CMOS-compatible $HfO_2$-based FE, especially 50% zirconium-doped $Hf_{0.5}Zr_{0.5}O_2$ (HZO) as well as antiferroelectric (AFE) $HfO_2$ material [4, 9] have attracted considerable interest. Ferroelectric capacitors (FeCaps) are useful for 1T-1C ferroelectric random-access memories (FeRAMs), [10] or as the gate stack in field effect transistors (FeFETs). [8, 11] AFE stacks are also attractive for FeFETs [4] due to their stability in addition to scalability and compatibility with existing semiconductor processes. The FE or AFE properties of HZO are governed by the stabilization of either polar orthorhombic (*o*-phase) or non-polar tetragonal (*t*-phase) structures.[12] The more stable antiferroelectricity [4, 13] is reflected in in a pinched shape or a double hysteresis loop of *P*–*E* characteristics. Rapid thermal annealing (RTA) is typically used to induce the formation of the desired polar phase. [14, 15] However, conventional high-temperature treatments for introducing high ratio of *o*-phase is not a viable option for BEOL integration due to thermal budget limitation. In addition, challenges such as high leakage, [16] extensive wake-up, [15] imprint, [17, 18] endurance degradation, [19, 20] and large switching voltage [21] restrict their application in low-power circuits, especially when thinking of integration with advanced node transistors.

Previous work by our group [22] demonstrated wake-up-free remanent polarization ($2P_r$) of 37 $\mu C/cm^2$ with enhanced endurance in HZO-based capacitors processed at 450 °C. However, high coercive fields ($E_C$) and asymmetric ferroelectric switching (different $E_{C+}$ and $E_{C-}$) still present critical obstacles to efficient device performance, mainly for energy efficient neuromorphic training applications. One promising strategy for improving low-voltage switching involves the use of superlattice $HfO_2$–$ZrO_2$ structures with equal fraction ratio (SL HZO), wherein alternating layers of $HfO_2$ and $ZrO_2$ with controlled sublayer thickness. Optimized superlattice structure can minimize defect accumulation at interfaces, [23] stabilize desired polar phases, [24, 25] and improve device endurance.[19, 23] However, achieving the optimal *o*-phase while maintaining BEOL-compatible thermal budgets remains challenging. Transition from AFE (or paraelectric) to FE phase has been demonstrated in doped $HfO_2$ systems [7, 12, 26, 27] by utilizing a custom electric-field cycling protocol (wake-up). Studies show that SL HZO structure with thicker sublayers tend to crystallize in the non-polar *t*-phase, requiring extensive wake-up pulsing to activate ferroelectricity such as wake-up pulsing of $10^3$,[19,28] $10^4$,[23,29] or $10^5$ [15] cycles.

In this work, we report on low-power switching, imprint-free SL HZO metal-FE-insulator-metal (MFIM) capacitors with record high $2P_r$ at 2 MV/cm field. Devices with varying sublayer thicknesses are fabricated and compared with MFIM capacitors grown with HZO solid solution, i.e. 1:1 ratio of $HfO_2$ and $ZrO_2$. Based on stack composition and annealing protocol, our devices show symmetric *P-E* hysteresis, $2P_r$ up to 76 $\mu C/cm^2$ at an operating voltage of ±2V, and pristine endurance exceeding $10^9$ cycles at 100 kHz frequency. Within ±1.8V sweep range, a $2P_r$ value of 53.4 $\mu C/cm^2$ ensures very high quality of the devices that can be integrated with 180 nm CMOS technology without need of additional circuitry. Furthermore, perfect linear analog control of polarization over a range of 40 $\mu C/cm^2$ and recovery from fatigue are achieved, positioning SL HZO FeCaps as promising candidates for future low-power non-volatile memory and neuromorphic computing applications.

## 2. Results and Discussion

### 2.1. Structural Characterization

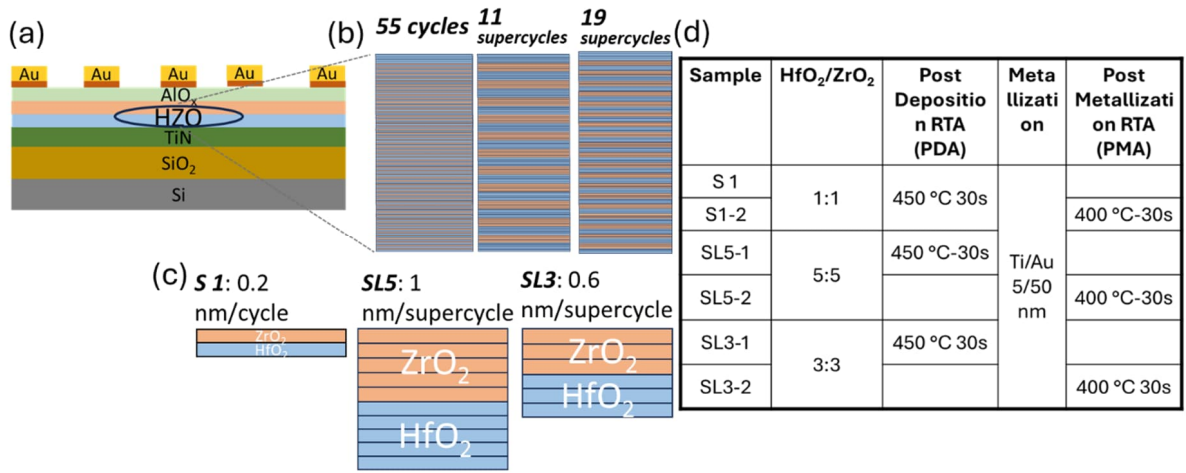

**Figure 1. (a)** Schematics of the MFM device stack in our experiments. **(b, c)** Various HZO stack grown in ALD of superlattice SL5 and SL3 in comparison with S1. **(d) Table** is an overview of thermal treatments carried out on all samples.

The MFIM material stack in this work is identical to our previous work [22] with various $HfO_2$-$ZrO_2$ sublayer thickness as shown in **Figure 1(a, b)**. SL5 refers to the HZO stack with 5:5 layers of $HfO_2$-$ZrO_2$ repeating for 11 supercycles resulting in 10 nm thickness. SL3 refers to the 3:3 layers of $HfO_2$-$ZrO_2$ repeating for 19 supercycles (**Figure 1(c))**. SL3 has nearly 0.5 Å higher thickness compared to the rest. The PDA samples, annealed at 450 °C before metallization are named SL5-1, SL3-1. The PMA samples, annealed at 400 °C after the metallization are named S1-2, SL5-2, SL3-2. (**Figure 1(d)**)

**Figure 2(a)** depicts the Grazing incidence x-ray diffraction (GIXRD) spectroscopy of the SL samples compared to the S1, S2 which have been PDA at 450 °C and 550 °C in previous work.[22] The diffraction peak position from the 2θ scans of SL(3,5)-2 match that of S1 and SL5-1. Hence, the reduced annealing temperature of 400 °C is found enough for the crystallization of the SL HZO into *o/t* phase dominant with no prominent diffraction peak of monoclinic (*m*) phase such as *m* (-111) and *m* (111) at 28.4° and 31.6°.[12] The high-resolution short range 2θ scans around 30° (inset of **Figure 2(a))**, shows that the reference diffraction peak of S2 shifted slightly to a smaller angle compared to the other samples. It aligns with previous reports that with an increase in RTA temperature higher percentage of *o*-phase can be obtained from the polycrystalline HZO.[15,16,22] Quantitative analysis of the contribution of *o*, *t*, *m* phase (supplementary **Figure S1**), show all samples are *t*-phase dominant and have less than 10% *m*-phase, which is similar to S1, however, superlattice SL5- (1,2) have an even higher composition of *t*-phase (approaching 0% *o*-phase) compared to the SL3 and S1, which accounts for the pristine state AFE behavior, as discussed in section 2.2. This is also in agreement with previous reports showing metastable *t*-phase is preferred in larger laminates. [19,30]

**Figure 2(b, c)** shows the TEM images of the SL5-2, where a wake-up pulsing of 1000 cycles has been applied. The material composition marked with yellow text is derived from the Energy Dispersive X-ray Spectroscopy (EDS) analysis. The TEM confirms the 10 nm SL HZO layer and 30 nm TiN layer in the MFIM stack, with a relatively well-defined interface. No distinct alternating sublayers of the superlattice $HfO_2$-$ZrO_2$ are observed attributed to grains formation with random domain orientations resulting from crystallization at lower temperature [23] of 400 °C. However, comparison of the areas marked with yellow and red contours in the enlarged image (**Figure 2(c)**) shows the epitaxial TiN grown with plasma-enhanced ALD (PEALD) has rather smaller lattice spacing than the superlattice HZO which has been designed to grow with thicker sublayer thickness. The STEM images from prior work [7] on Si doped $HfO_2$ revealed that the relaxation to *t*-phase at the $HfO_2$-TiN interface becomes less pronounced after wake-up pulsing. A recent work,[25] using the high-resolution STEM, has shown that SL HZO with ratio of 3:3 similar to SL3 in this work, exhibits the most stable ferroelectricity due to the highest portion of *o*-phase. In the following sections, we describe the electrical measurements to understand the distribution of FE and AFE phases at pristine, woken-up, fatigued and recovered state of the SL HZO together with rationale for lower $E_C$ and imprint-free switching.

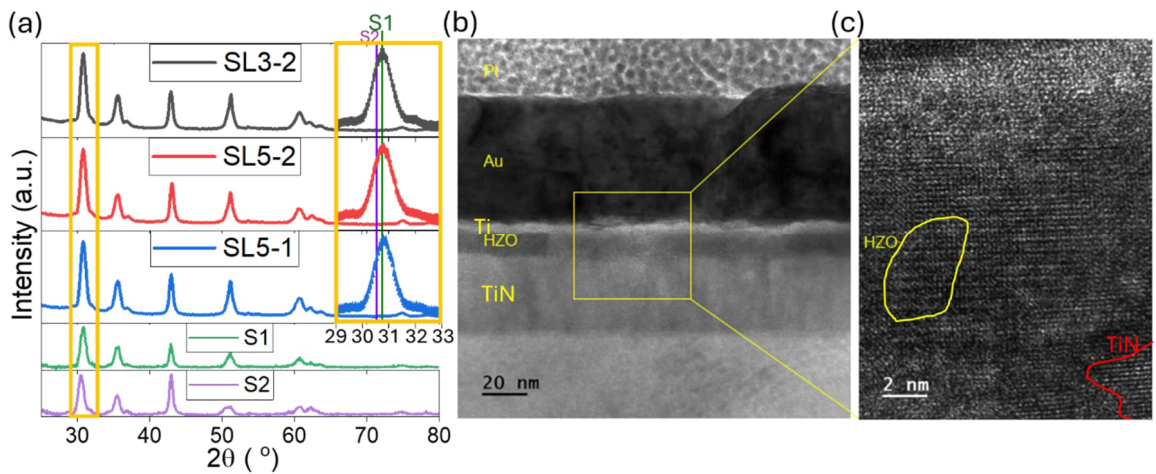

**Figure 2.** (a) Grazing incident x-ray diffraction (GIXRD) full range 2θ scan from 25° to 80° of SL(3,5)-2, SL5-1 with reference S1 and S2 from previous work[22] at the bottom. Inset is the short range high resolution 2θ scan from 29° to 33° of SL(3,5)-2, SL5-1, the reference diffraction of S1 and S2 are added as vertical lines. **(b, c)** Transmission electron microscopy (TEM) images of the stack showing superlattice growth of SL HZO and epitaxial growth of TiN of the SL5-2 which has been PMA at 400 °C.

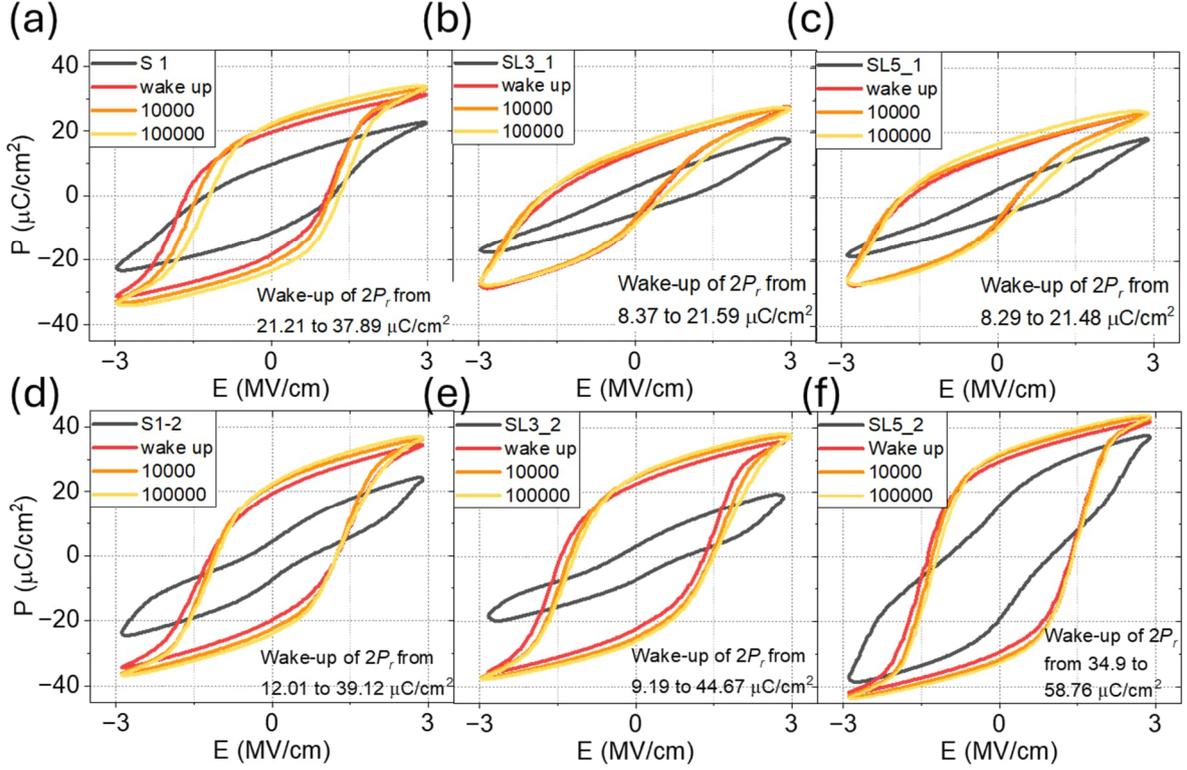

**Figure 3.** Polarization vs. electric field (*P-E*) measured in DHM mode at 3 V, 1k Hz frequency of samples **(a)** S1, **(b)** SL3-1, **(c)** SL5-1, **(d)** S1-2, **(e)** SL3-2, **(f)** SL5-2. Black curve represents device pristine response before any pulsing, red curve is the woken-up *P-E* hysteresis after $10^3$ pulsing cycles, orange is after $10^4$ pulsing, yellow is after $10^5$ pulsing showing initial large improvement of $P_r$ followed by stabilization.

## 2.2. Electrical Characterization
### 2.2.1. Wake-up from pristine state

Wake-up, i.e. opening of the pinched hysteresis in the pristine device through electric field cycling is a prerequisite not only in SL structures but also in many solid solution films such as Sr doped $HfO_2$,[7] Rd doped $HfO_2$,[31] PEALD grown HZO,[15] HZO PMA at low temperature (400 °C),[32] and at high temperature (700 °C).[33] In the present work, a rectangular pulse of ±3 V at 1 kHz was applied as wake-up protocol for all samples. The wake-up pulsing scheme is chosen based on the wake-up percentage, leakage contribution, and time-energy balance by comparing the field cycling with various frequencies (1k, 10k and 100k Hz). A dynamic hysteresis measurement (DHM) with triangular pulse was applied in pristine state and every decade to monitor the wake-up process.

**Figure 3** shows the hysteresis measured with 3 V, 1 kHz in DHM mode for all samples. As shown in **Figure 3a** black curves, solid solution sample S1 exhibits FE behavior already in pristine state, which gives a wake-up free $2P_r$ of 20 µC/cm$^2$. With alternating thicker sublayer, same PDA treatment as S1, SL (3, 5)-1 (black curves in **Figure 3(b, c)**), show no FE behavior in its pristine state, and a rather symmetric AFE behavior for S1-2 and SL5-2, which have been PMA treated at lower temperature 400 °C. Although all PDA treated samples have asymmetric hysteresis shown as uneven $E_C\pm$ and $P_r\pm$, the pristine P-E hysteresis of SL (3, 5)-1 are more pinched, manifested in small $2P_r$ values. This can be attributed to the following factors: 1) crystallization of metastable $t$-phase is preferred for large laminates as discussed in GIXRD results in Section 2.1, resulting in AFE behavior for SL5 and SL3, which have been grown with increasing thickness of sublayer. 2) Without a post-metallization annealing, there exists a resistive gap layer at the ALD grown FE stack and the top electrode (TE) interface. This layer contributes to a voltage drop across the TE interface, causing the devices to switch at a higher voltage in the negative voltage direction. Additionally, 3) the dielectric capping layer of AlOx,[34] resistive gap layer at the interface and thicker HfO$_2$ or ZrO$_2$ layers inside the stack contributes to depolarization field ($E_{dep}$) and inhomogeneous charge distribution at pristine state. The first two factors account for the sizable $\Delta E_C\pm$ observed in SL(3,5)-1, while the latter one may give rise to the pinched hysteresis in SL samples. Moreover 4) in SL structures, a thicker HfO$_2$ (from 1 to 5 monolayer) layer at the bottom may shield the influence of in-plane stress from the PEALD grown TiN, which has been considered crucial for achieving high $P_r$.[35]

Upon pulsing with an electric field ($E$), the pinched AFE hysteresis gets transformed into a more open (squared shape) P-E with higher $P_r\pm$ as shown in **Figure 3**. The transition to FE hysteresis after wake-up field cycling, is due to either a transition from the $t$-phase to the polar $o$-phase,[12,36] or trap-state filling triggered by the charge redistribution during wake-up pulsing[7] acting as screening charges and destabilizing $E_{dep}$ or a combination of both effects. After wake-up pulsing for 1 s (1000 cycles), all samples have a certain degree of increase in $2P_r$. The wake-up (1 s) percentage of S1, SL3-1, SL5-1, S1-2, SL3-2 and SL5-2 are 79%, 158%, 159%, 225%, 386% and 68% successively. There is only a slight increase (less than 15%) of $2P_r$ upon further pulsing shown as the yellow color curves (wake-up 100 s) in **Figure 3**, which can be attributed to the increasing leakage current contribution[37] caused by defect (O vacancy) creation.[7] The 1s (1000 cycles) wake-up pulsing shows no contribution to the static leakage of the device as shown in the static *I-V*s in supplementary **Figure S2**, therefore chosen as standard wake-up pulsing protocol.

After wake-up pulsing, a pronounced imprint effect remained in PDA-treated samples, as shown in **Figure 3(a, b, c)**, where the $\Delta E_{C}\pm = |E_{C-}| - |E_{C+}|$ is 1.2 MV/cm for SL(5, 3)-1, and 0.6 MV/cm in S1. The imprint, *i.e.* non-zero $\Delta E_C\pm$ or a variance of $P_r\pm$ is caused by pronounce depolarization field. [34,38] Larger imprint effect in PDA treated SL samples can be attributed to the combined effect of the less *o*-phase formation and transition resulting in smaller $E_{C+}$ compared to $E_{C-}$ and voltage drop at the top interface leading to less electric field across the HZO layer, as also observed in previous works.[34,38,39]

The large $\Delta E_C\pm$ in PDA-treated samples is reduced in PMA-treated samples to 0.1 MV/cm for the woken-up S1-2, SL3-2 and becomes fully symmetric in the woken-up SL5-2. The $2E_C$ of the woken-up S1-2 and SL(5, 3)-2 are 2.4 and 2.7 MV/cm with improved switching symmetry and higher $2P_r$. The woken-up $2P_r$ of all PMA-treated samples are found to be higher than that of PDA-treated samples with an increase of 3%, 106% and 173% for 1:1, 3:3 and 5:5 stack. Benefiting from the imprint-free feature (symmetric $E_C\pm$) arising from PMA treatment, a high $2P_r$ is achieved with reduced $E_C$. Improvement in $2P_r$ found in 5:5 stack can be attributed to the emergence and coexistence of antiferroelectricity (more *t*-phase formation) already in pristine state. The presence of AFE fractions in SL structure can stabilize the intermediate *t*-phase [40] leading to low $E_c$.[19] However, more experiments are needed to support the idea that *t*-phase can stabilize better in the 5:5 stack.

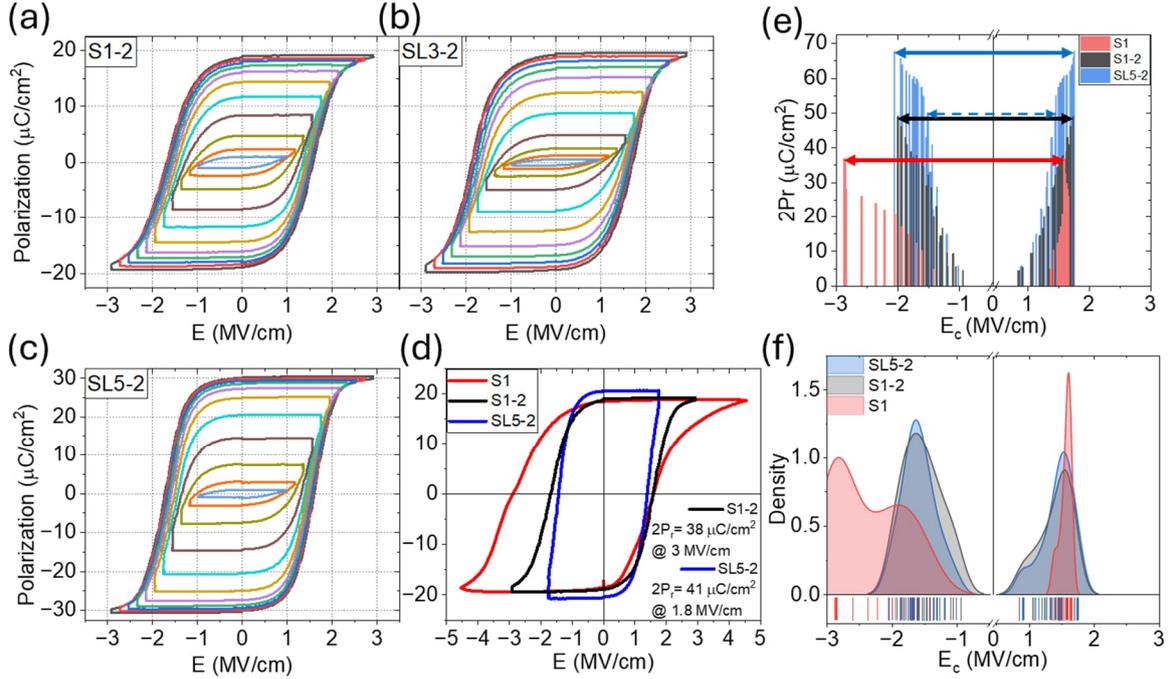

**Figure 4.** Post 1s wake-up pulsing, *P-E* hysteresis loops measured using PUND pulsing scheme of PMA-treated samples with **(a)** 1:1, **(b)** 3:3, **(c)** 5:5 stacking configurations. The measurements were performed with descending electric fields from 3 to 1 MV/cm applied. **(d)** Comparison of *P-E* hysteresis, which give $2P_r$ around 40 µC/cm². **(e)** Evolution of switching polarization ($2P_r$) and corresponding coercive field $E_c\pm$ with various switching fields measured up to each sample's maximum tolerance before HBD. **(f)** Distribution of $E_c\pm$ values for S1 (from previous work) and S1-2, SL5-2 in this work over multiple switching cycles demonstrating improvement in $E_c$ symmetry.

### 2.2.2. Improved FE characteristics and Imprint-free operation

PUND (positive-up-negative-down) measurements subtract the dielectric contribution from the overall charge switching, revealing accurate FE characteristics such as $P_r\pm$, $E_c\pm$. The *P-E* hysteresis in **Figure 4(a, b, c)** are measured with descending pulsing amplitude from ±3 to ±1 MV/cm for the PMA-treated samples after wake-up cycles. At 3 MV/cm, the switching polarization ($2P_r$) of S1-2 and SL3-2 reaches a similar value around 40 µC/cm², which is substantially lower than that of SL5-2 (60 µC/cm²). As the applied electric field decreases, the $P_r$ decreases. S1-2 exhibits a broader switching current peak resulting a slightly broader *P-E* hysteresis (**Figure 4(a)**). In contrast, SL 5-2 shows a sharper switching current and a more square-shaped hysteresis (**Figure 4(c)**).

To reach a switching polarization ($2P_r$) of about 40 µC/cm$^2$, the PDA-treated 1:1 stack (S1) requires an electric field of 4.6 MV/cm, which can be reduced by PMA treatment to 3 MV/cm for S1-2 (**Figure 4(d)**). Further improvement in the write voltage efficiency is observed in SL5-2. With only 1.8 MV/cm applied (only 40% of the field required for S1), SL5-2 can already give a $2P_r$ of 41 µC/cm$^2$. **Figure 4(e)** summarizes the measured $2P_r$ (bar height) and corresponding coercive fields ($E_c\pm$: bar position) from each PUND loops of S1, S1-2 and SL5-2. Before hard break down (HBD), S1, S1-2 and SL5-2 tolerate 4.6, 4.2 and 4.2 MV/cm applied field respectively, yielding maximum $2P_r$ of 37, 49 and 68 µC/cm$^2$ respectively. The general trend is higher $2P_r$ requires a higher applied voltage and leads to a broader memory window ($2E_c=|E_{C-}| + |E_{C+}|$). However, significant improvement in $2E_c$ as well the symmetry of $E_c$ is observed for PMA samples (**Figure 4(e)**). Additionally, the overall $E_c\pm$ distribution of SL5-2 is much narrower compared to S1 over multiple switching cycles showing better reliability (**Figure 4(f)**). As shown in **Figure 4**, all PMA samples are lesser imprint than PDA-treated samples, among which SL5-2 has identical $E_{c-}$ and $E_{c+}$ of 1.4 MV/cm, while SL3-2, S1-2 has a $\Delta E_c\pm = 0.1$ MV/cm. Among 3 PMA-treated samples, 5:5 stack shows the best performance in features like high $P_r$, low switching voltage and symmetric hysteresis. These merits directly translate to reduced write/erase voltages and improved power efficiency in device operation.

In **Figure 5**, we show that even higher polarization switching is attainable by reducing the junction area. **Figure 5(b)** shows sample SL5-2 can reach up to 90 µC/cm$^2$ at 3 MV/cm and 76 µC/cm$^2$ at 2 MV/cm in a 0.01 mm$^2$ junction, which is significantly higher compared to the 0.04 mm$^2$ junctions (**Figure 5 (a)**) and one of the best reported values in literature (**Figure 5 (c, d)**). A thorough junction area dependence of polarization is an ongoing study and will be reported in future works. The data point from this work in benchmark **Figure 5 (c, d)** are from the PUND values of SL5-2 and endurance shown in section 2.2.4.

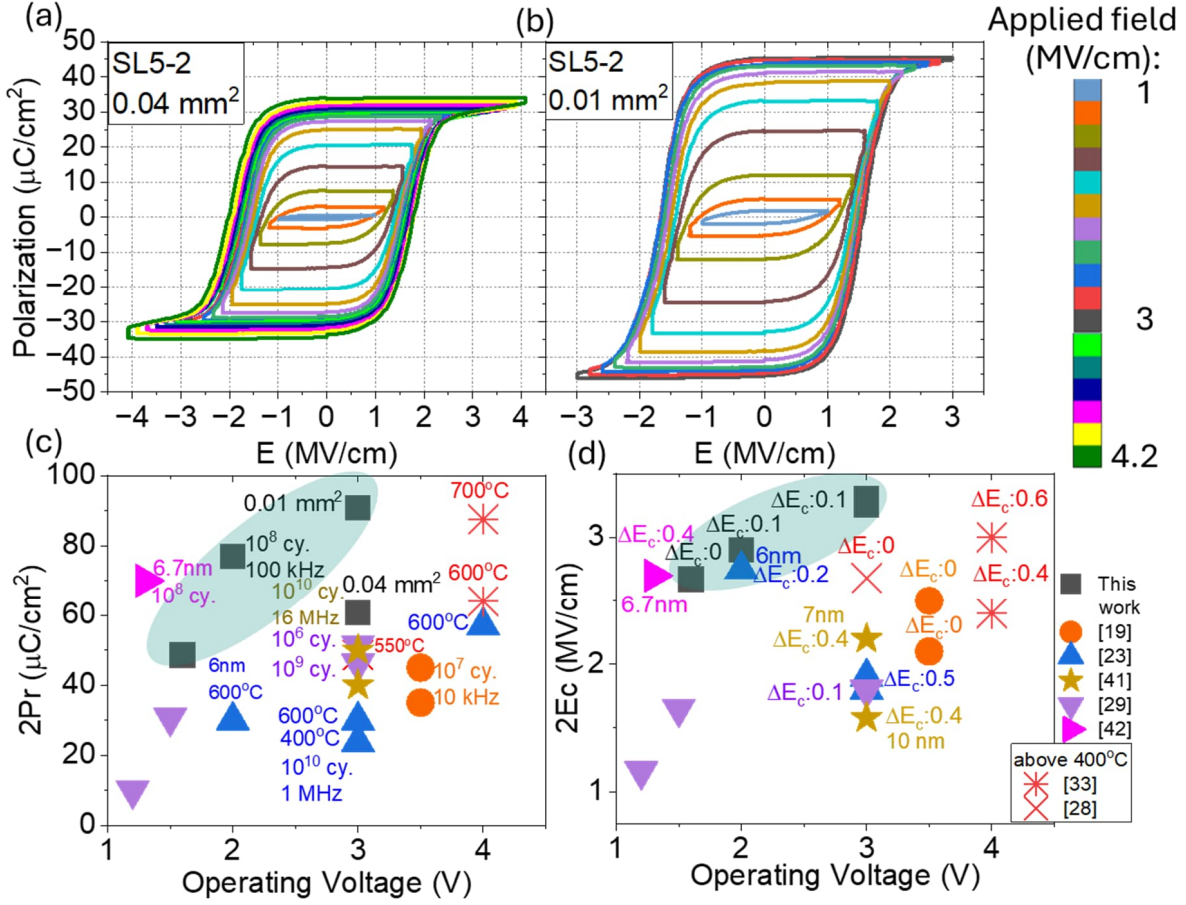

**Figure 5.** Post wake-up, *P-E* hysteresis loops measured using PUND pulsing scheme of PMA-treated samples SL5-2. **(a)** The *P-E* loops measured on 0.04 mm² junction area with electric fields ranging from 1 to 4.2 MV/cm compared with **(b)** 0.01 mm² junctions where significantly higher $P_r$ is obtained within 3 MV/cm. Benchmark of **(c)** switching polarization $2P_r$ value and **(d)** memory window $2E_c$ as a function of operating voltage for MFM HZO devices reported in literature. The results reported in this work (in green circles) show excellent performance of our SL stack, promising for low-power NVM and neuromorphic application.

### 2.2.3. Analog control

Due to their improved $P_r$ and switching symmetry, the PMA-treated samples are tested for analog control, that is essential for in-memory computing (IMC). As shown in **Figure 6(a)**, after wake-up, the $P_r$ can be modulated almost linearly over a broad range by increasing amplitude of applied voltage. Overall, sample SL5-2 shows the highest response reaching a higher $2P_r$ up to 60 μC/cm², whereas the others reach around 40 μC/cm² within 3 MV/cm applied field. Remarkably, within 1-2 V pulsing range, the SL5-2 achieves a change of $2P_r$ of 40 μC/cm² with almost perfect linearity for both potentiation and depression sides, shown by the green dashed line in **Figure 6**. It is reported recently that a $2P_r$ of 40 μC/cm² can produce 16 reliable analog states in HZO- based FeCaps and can work perfectly as a 4-bit memory.[43] In SL devices, domain switching majorly happens at low field that leads to rapid change in $P_r$. At higher field, only a few domains switch, resulting in a near saturation state and a deviation from linearity. During the depression, the saturation appears between 3 V to 2.2 V, which is slightly less than that of potentiation. For high accuracy IMC, a linear and symmetric weight update is essential and a low voltage (1-2V) operation of SL-HZO can provide these highly linear properties. Superior polarization tunability of SL HZO at lower V makes them attractive as tunable synaptic weights in IMC hardware.

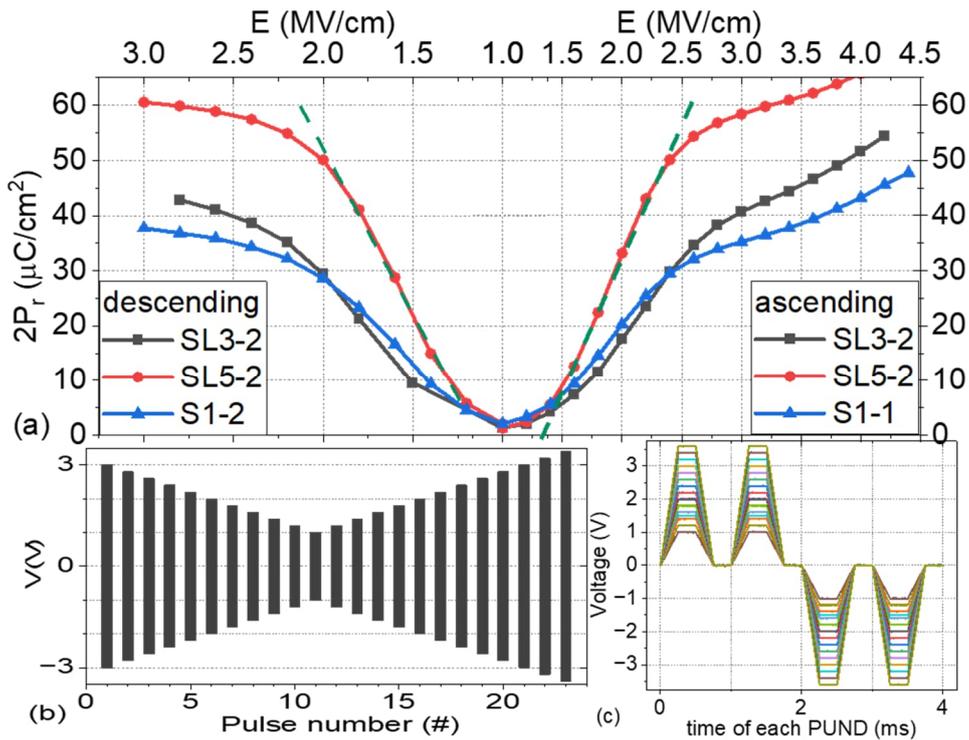

**Figure 6.** Post wake-up, modulation of remanent polarization by pulsing amplitude. (a) depression and potentiation of $2P_r$ measured on one junction without break down of the samples SL3-2 in black curve, SL5-2 in red curve and S1-2 in blue curve, in response to voltage pulse with descending amplitude from 3 to 1 V (0.2 V/step) and ascending amplitude from 1 V to 4.2 V illustrated in (b). Each PUND pulsing scheme (c).

**2.2.4. Endurance Characteristics**

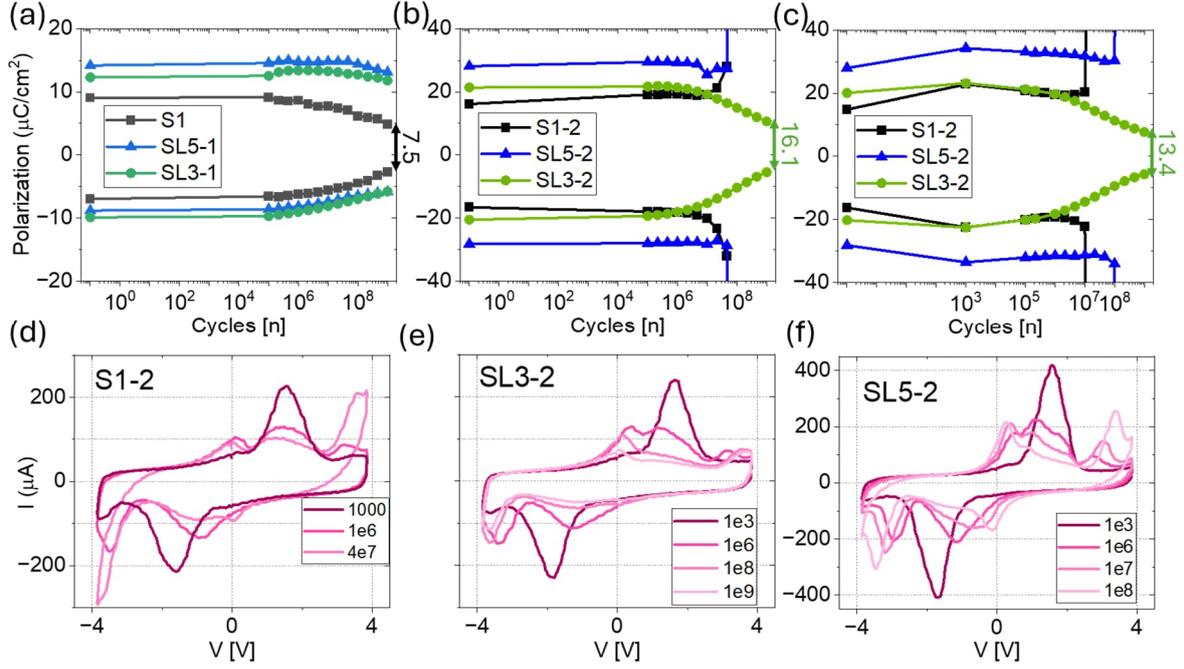

**Figure 7.** Endurance test with trapezoid pulsing of ±3 V, 100k Hz of samples with stack 1:1 in black curves, 5:5 in blue curves and 3:3 in green curves. **(a)** Samples PDA-treated at 450 °C (monitor DHM at 4 V for SLs, 3.5 V for 1:1) and **(b, c)** PMA-treated at 400 °C. **(b)** without wake-up **(c)** post wake-up pulses. Evolution of *I-V* loops in DHM mode during the endurance test for **(d)** S1-2, **(e)** SL3-2, and **(f)** SL5-2. Results show a gradual increase of leakage for hard breakdown in 1:1 stack, increase of leakage until $10^7$ cycles followed by only fatigue to dielectric behavior gradually until soft breakdown of 3:3 stack and a transition to AFE behavior before hard breakdown of 5:5 stack.

**Figure 7(a)** compares the endurance of PDA-treated samples. The SLs are monitored with 4 V for 0.04 mm$^2$ junctions whereas S1 has been monitored at 3.5 V for 0.25 mm$^2$ junction. The SLs can survive a higher voltage amplitude at a smaller junction area due to less leakage and improved device reliability although the imprint manifested as uneven $P_r\pm$ in **Figure 7(a)** and $E_c\pm$ in **Figure 3(b, c)**. After $10^9$ cycles of pulsing, the $P_r$ reduced to 47% of the pristine value for S1, while it remained at 79% and 82% of the pristine $P_r$ in SL3-1 and SL5-1. The evolution of hysteresis shapes is shown in supplementary **Figure S3**. The degradation of the $P_r\pm$ can be attributed to the continuous imprint effect upon pulsing rather than effect of the switching current degradation (**Figure S3(a)**)**.** The endurance pulsing for SL(3, 5)-1 results in a slow wake-up process rather than inducing fatigue. Therefore, the SL structure with higher sublayer thickness is more promising for a better stability performance, and high endurance.

As illustrated in Section 2.2.2, imprint-free FE hysteresis can be achieved for both SL and S1 samples by PMA treatment combined with wake-up pulsing. Although SL(3, 5)-2 have AFE characteristics in the pristine state but can be transformed into FE states upon wake-up pulsing. In **Figure 7(b, c)**, we show the endurance of the PMA-treated samples without and with wake-up pulsing. SL3-2 exhibits the longest endurance without breakdown until $10^9$ cycles with a certain degree of fatigue. Static *I-V*s in supplementary **Figure S2(c, d)** shows that the leakage contribution of SL3-2 remains lower than that of SL5-2 even after one extra decade of pulsing. On the other hand, although S1-2 and SL 5-2 breaks-down earlier (after $10^7$ and $10^8$ cycles, respectively), it shows better stability of $P_r\pm$ upon pulsing.

Dynamic *I-V* loops during endurance tests of PMA-treated samples are shown in **Figure 7(d, e, f)**. **Figure 7(b, c)** shows that 1:1 stack breaks down earliest among all samples due to continuously increasing leakage contribution upon pulsing, as shown in **Figure 7(d)**. Without wake-up pulsing, the endurance of SL3-2 lasts until $10^9$ cycles. There is a slow degradation of the $2P_r$ of nearly 17% until $10^7$ cycles. Upon further pulsing until $10^9$ cycles, 38% ($2P_r$: 16.1 µC/cm$^2$) of the pristine $2P_r$ remains. When using dedicated wake-up, faster fatigue appears in SL3-2 (25% degradation until $10^7$ cycles) followed by 33% ($2P_r$: 13.4 µC/cm$^2$) remanence at $10^9$ cycles, which is slightly faster compared to the one without wake-up pulsing. This fatigue results from the combined effect of switching current degradation and imprint upon pulsing as can be seen from the monitoring *I-V*s in **Figure 7(e)**. The dynamic leakage for the positive electric field is well suppressed in SL3-2 while for the negative electric field, the leakage current increases gradually until $10^7$ cycles and it starts to decrease upon further pulsing, showing reduced FE domain switching upon field pulsing beyond $10^7$ cycles. For SL5-2, there is only 5% degradation of woken-up $2P_r$ upon pulsing $10^8$ cycles i.e. still 114% of pristine $2P_r$, and 0.4% degradation of pristine $2P_r$ before its HBD. Notably, the evolution *I-V*s of SL5-2 during endurance test, as shown in **Figure 7(f)**, is rather different. An AFE characteristic evolved upon pulsing (a transition from woken-up FE to AFE). Without pre-wake-up pulsing, SL5-2 exhibits only 7% wake-up of $2P_r$, followed by abrupt breakdown upon $4*10^7$ cycles which can be improved in the woken-up SL5-2. The woken-up SL5-2 shows a gradual transition from FE to AFE states upon pulsing with 100 kHz, before the breakdown, it endures until $10^8$ cycles (**Figure 7(c)**). Hence such phase transition between AFE to FE state at a certain condition, achieved by electric field cycling with ±3 V, 100k Hz here, is reversible in nature in SL5-2 rather than irreversible phase transition from *t* to *o*.[30,44,45] Stack of 5:5 (SL5-2) can either be crystallized in *t*-phase or stabilized within *o*-phase as in-plane oriented domain gives rise to the AFE phase at pristine state[36] and transforms to FE phase when destabilization happens due to defect (dislocation) generation upon wake-up pulsing and subsequently switching back to AFE behavior due to the redistribution of defects while pulsing with 100k Hz during endurance test.

This shows different fatigue (or wake-up) mechanisms are present for 1:1, 3:3 stacks and 5:5 stacks. SL5-2 offers a larger $2P_r$ with the best stability of $P_r\pm$ and $E_c\pm$ upon pulsing with less endurance. SL3-2 gives the best endurance, however with fatigue of $P_r$ and imprinted $E_c$. Section 2.2.5 discusses the possibility of recovery, both from fatigue and imprint for SL3-2 and reversibility from AFE- to FE behavior for SL5-2 before breakdown.

### 2.2.5. Recovery from fatigue

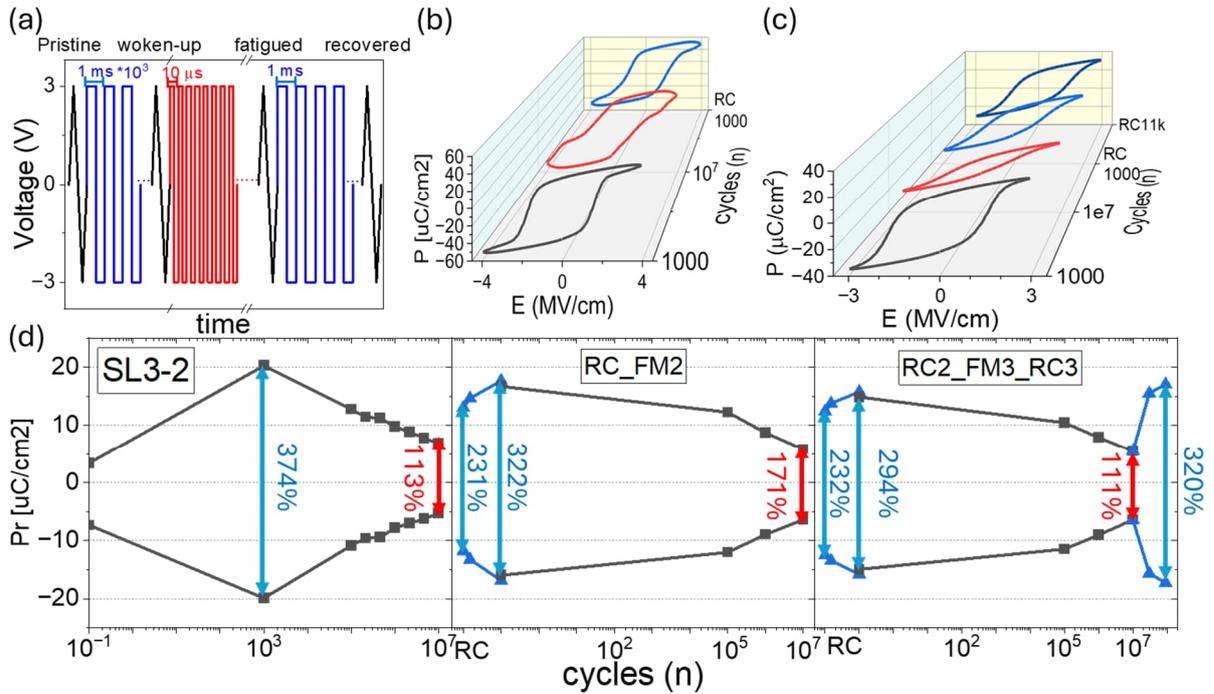

**Figure 8. (a)** Pulsing schemes employed for wake-up, fatigue and recovery. Recovery from fatigue measurements performed on SL5-2 showing reversible transition of from woken-up FE in black curves to fatigued antiferroelectric in red curves and recover (RC) to FE in blue curves with 100% $2P_r$ in woken-up state within 1 s of wake-up pulsing. **(b)** *P-E* hysteresis in accordance with cycle number of SL5-2, monitoring DHM was carried out at 4 V, 1 kHz. **(c)** *P-E* hysteresis of SL3-2, monitoring DHM was carried out at 3V, 1 kHz, recovery pulsing for 11 s (11k cycles), the first recovery process in (d). (d) Recovery from fatigue after each time that switching polarization has fatigued to approximately 12 μC/cm² (3 times) performed on SL3-2, showing evolution of $P_r$ as a function of cycle number. $2P_r$ of above 30 μC/cm$^2$ can be achieved reproducibly after each recovery cycle.

Recovery from fatigue has been considered an effective way to achieve a high endurance non-volatile memory device and enable its application in DNN online training. Various recovery pulsing schemes have been reported for HZO capacitors,[20,22,46] in 1T1C FeRAM,[10] FeFETs,[11] and in SL HZO.[28] Besides, recovery from imprint FE hysteresis (either from pristine state or from high temperature) have been demonstrated by field cycling for $10^4$ cycles or sweeping with higher electric field[17] and by conducting the measurement in cryogenic environment.[47] Taking the cost of energy into consideration, an ideal recovery method should balance the recovery field magnitude, and the number of pulsing cycles required for certain recover percentage to achieve the highest recover efficiency.

As shown in Figure 8(a), the recovery in this work uses the same pulsing scheme as wake-up pulsing ±3 V, 1kHz. Depending on the degree of fatigue $2P_r$, recovery (RC) pulsing with number of cycles from 1000 (1 s) to 200k (200 s) has been carried out. SL5-2, showing slower fatigue speed (8% degradation of woken-up $2P_r$ until pulsing $10^7$ cycles) and AFE characteristic at fatigued state (post $10^7$ cycles), can be recovered to 107% ($2P_r$: 76.2 μC/cm$^2$) of the woken-up state ($2P_r$: 70.6 μC/cm$^2$) with 1 s of RC pulsing as shown in Figure 8(b), where in SL3-2 after $10^7$ pulsing cycles, 11k cycles of RC pulsing (11 s) can achieve a recovery of 86% shown in Figure 8 (c). Similarly, the degradation in $2E_c$ of SL3-2 to 48% can be recovered to 80% with better symmetry ($\Delta E_c \pm$ 0.1 MV/cm) after 11k RC pulsing. Without as much degradation as in SL3-2, the $E_c\pm$ of SL5-2 is quite stable upon pulsing, after recovery, a slight (9%) reduction of $2E_c$ (mainly the $E_c$-) is observed. The fast reversible transition between AFE to FE characteristics confirms that instead of irreversible phase change happening, SL5-2 has $o$ phase stabilized in in-plane domain and would be destabilized [36] due to defects generation caused by the 1k Hz field pulsing.

For the extreme case in SL3-2, *i.e.,* after $10^9$ pulsing, a 200 s RC pulsing can recover 60% ($2P_r$: 27.85 μC/cm$^2$) of the woken-up $2P_r$ as shown in Figure S4. Therefore, recovery from dielectric or strongly fatigued phase to FE is more difficult and requires more energy for the RC. The fast degradation and longer recovery pulsing can be attributed to the formation of switching region and non-switching region when high frequency and low field applied, which leads to the defects diffusion due to the inhomogeneous charge density at fatigued state.[28,48,49] Such mechanism agrees with the fact that diffusion of large number of oxygen vacancy related defects at the FE/metal interface causes HBD which is the case of S1, and a smaller number of defects in SL samples diffuse to trap level that need higher RC energy, which is the case of SL3 and finally, further less amount of defect in SL5 results in better stability and fast recovery upon pulsing.

**Table 1** presents the benchmarking of emerging HZO devices in the recent 5 years. In general, nearly all HZO devices have wake-up effects. Nonetheless, there is significant improvement on the endurance of the device after superlattice structure or dielectric layers have been involved. However, if taking the pulsing frequency of endurance test into consideration, all samples in this work were tested with 100 kHz, which gives more convincing results and expected to endure longer cycles, if pulsing with higher frequency. By only 1 s (1000 cycles) of wake-up pulsing with ±3 V 1 kHz, the post metallized SL HZO in this work has shown one of the most competitive performances at low operating voltage, FE polarization switching symmetry, device-to-device performance reliability together with recoverable fatigue and long endurance. Progressing forward, thinner FE films with high polarization and endurance would be one direction for even lower power operating devices, as shown for perovskite ferroelectric devices.[50,51] Also, by proper engineering of the depolarizing field, fabricating FE leaky-integrate-and-fire neuron[52] would be another target for a fully-integrate FE neural network fabricated on CMOS back-end-of-line.

**Table 1.** Table benchmarking the present work with the state-of-the art results from FE HZO capacitors in terms of thermal treatment, wake-up pulsing, voltage requirement, symmetry of $E_c\pm$ and endurance properties.

| No | Device architecture (BE/HZO/TE) | PMA Temperature (°C) | Applied Electric field (MV/cm) | Wake up cycles | $P_r$ (μC/cm$^2$) | MW, $E_c\pm$ (MV/cm) | Endurance (cycles) | Recovery (2$P_r$/ 2$P_r$ wake-up) | Reference |
|---|---|---|---|---|---|---|---|---|---|
| 1 | TiN/HZO/TiN/Ti /Pt (7 nm HZO) | 400 PMA 1h | 3 1k Hz | 3 V 10$^4$ cycles 10 s | 15 | 2 (Ec:1) RT Δ Ec: 0.4 V 200 for 12 h | 2*10$^6$ 1k Hz | recovery from imprint by 10$^4$ pulsing | [17] |
| 2 | TiN/HZO/TiN (15 nm HZO) | 450 PMA 30s | 4 100 Hz | 4 V 100 Hz 10$^4$ cycles 100 s | 17.8 (HO) 11.7 (LO) | 3.5 ΔEc: 0.6 RT ΔEc: 0.9 (LO) ΔEc: 1.5 (HO) | 5*10$^6$ (HO) 10$^{11}$ (LO puls O plasma treatment) | | [18] |
| 3 | TiN/HZO/WN$_x$ /Ru (10 nm HZO) | 400 PMA 1 h | 4 1 kHz | 3V, 100 Hz, 10$^3$cycles | 17.5 | < 3 Ec+: 1.65 Ec-: -1 | >10$^9$ 2V, 1 MHz | 82% | [20] |
| 4 | TiN/SL HZO (1.25:1.25 nm)/TiN | 350-600 PMA | 3 | 3 V 10$^4$ cycles | 11.5 (350) 15.8 (600) | 2 Ec+: 1.2 Ec-: -0.8 | 2*10$^{10}$ (350) 10$^{11}$ (600) 3 V, 1 MHz | - | [23] |
| 5 | TiN/SL HZO (here is 0.5:0.5 nm)/TiN/Ti/Pt | 400 PMA 60s | 3.5 1k Hz | 3 V 10$^3$ cycles 1s | 20 | 2.5, Ec±: 1.25 2.1 (2.5 nm) | <10$^7$ 2.75 V, 10 kHz | | [19] |
| 6 | TaN/SL HZO(1:1 to 15:15) /TaN | 550 PMA 30s | 3 10k Hz | 3 V 10$^3$ cycles 0.1s | 22 (5:5) | 3 Ec+: 1.6 Ec-: 1.4 | 3*10$^7$ 2V, 1 MHz | >100% | [28] |
| 7 | TiN/TiO2/SL HZO (0.625:0.625 nm)/TiN/Al | 350-600 PMA 30s | 3 1k Hz | 3 V 10$^4$ cycles 10s | 25.7(400) 20 (350) | 1.8 Ec±: 0.9 | 10$^6$ PMA400 10$^9$ PMA350 1 MHz | | [29] |
| 9 | TiN/SL HZO (3:3)/AlO$_x$/Ti/Au | | | 3 V 10$^3$ cycles 1s | 20 @ 3 V | 3.5 ΔEc: 0.2 | >10$^9$ 3 V, 100k Hz | 93% | This work |
| 10 | TiN/SL 0.04 mm$^2$ HZO (5:5)/AlO$_x$/Ti/Au 0.01 mm$^2$ | 400 PMA 30s | 3 1k Hz | 3 V 10$^3$ cycles 1s | 29 @ 3 V 20 @ 1.8 V 45 @ 3 V 38 @ 2 V | 2.8 ΔEc: 0.04 @ 1.8 V 2.9 ΔEc: 0.1 @ 2 V | >10$^8$ 3 V, 100k Hz | >100% | This work |

## 4. Conclusions

ALD grown superlattice $Hf_{0.5}Zr_{0.5}O_2$ with ratio of 3:3 and 5:5 is studied in this work. Post metallized rapid thermal annealing at 400 °C resulted in improved imprint-free polarization switching compared to the post-deposition annealed ones. Antiferroelectric behavior with very small switching polarization values were found in the pristine state of the SLs. After 1000 (1 s) wake-up pulsing cycles, transition to ferroelectric phase occurs. The results suggest that the electrical behavior of the SLs cannot be fully attributed to the phase transition from *t* to *o*. High remanent polarization results from the combined effect of the phase transition and domain depinning during wake-up pulsing. Similarly, the fatigue of polarization switching upon pulsing results from a combined contribution of domain pinning, reversible phase transition and charge trapping site formation upon long pulsing cycles. The performance of SL HZO in this work is beyond state-of- art devices considering the criteria of power consumption (including thermal budget, required electric field for wake-up and switching), and FE performance (including high $P_r$, good symmetricity, long endurance with possibility of recovery). Therein, the SL stack with 5:5 shows better stability due to the transition to stable AFE upon pulsing. Integration of FE and AFE is expected to exhibit more reliable and versatile functionalities for future applications in memory and neuromorphic hardware.

## 5. Experimental Section/Methods

**Fabrication methods:** The Metal-FE-Insulator-Metal capacitors, with structures shown in **Figure 1(a)**, were fabricated using Atomic Layer Deposition (ALD). Detailed ALD recipe can be found in [22]. In this work, superlattice HZO are obtained by alternating growth of the $HfO_2$ and $ZrO_2$ with ratio of 5:5 and 3:3 meanwhile maintaining the whole thickness to 10 nm. After the film deposition, the sample was cleaved into one-quarter of a wafer. The capacitor devices were completed by evaporating Ti/Au (5/50 nm) as top electrode through electron beam evaporation. PDA and PMA thermal treatments were carried out similarly for 30 seconds under nitrogen atmosphere at annealing temperatures of 450 °C, 400 °C respectively (samples name ended with -1 refers to PDA and -2 to PMA treatment respectively) as described in the Table in **Figure 1**.

**Structural Characterization**: Structural characterizations of the samples were carried out by GIXRD in Rigaku SmartLab diffractometer operating at 45 kV, 150 mA, with a Cu rotating anode. The incident angle ω was set at 0.35°, where the angle was at half maximum reflectivity intensity. TEM thin film is prepared with focused ion beam microscope (FIBSEM, Zeiss Crossbeam 540) and characterized with a (scanning) TEM (S/TEM, JEOL JEM-F200) equipped with an electron dispersive spectroscopy system (EDS, JEOL).

**Electrical Characterization**: Electrical characterizations of the samples were carried out by applying voltage between the bottom TiN and the top gold electrode. For ferroelectric polarization – electric field (*P–E*) hysteresis, endurance and retention measurements, ferroelectric material tester AixACCT 2000E was used. Current – voltage (*I-V*) characteristics were measured with a semiconductor parameter analyzer Keysight 1500A. Ferroelectric characteristics were measured on 200 x 200 $\mu m^2$ and 100 x 100 $\mu m^2$ square junctions with 1 kHz of triangle voltage sweep in DHM mode and 250 µs of pulse width of trapezoid pulse in PUND mode. For endurance measurement, a rectangular voltage pulsing of 3 V, 100 kHz frequency was applied followed by 3 DHM measurements per decade to monitor the $P_r$ value of the sample. All measurements were carried out at room temperature and under ambient conditions.

**Conflicts of interest**

There are no conflicts to declare.

**Data availability**

The data supporting the conclusions of this article is available upon reasonable request from the corresponding author. Some data have been included as part of the Supplementary Information.

**Ethical Statement**

There are no special ethical concerns to consider.

**Acknowledgements**

The authors acknowledge financial support from Research Council of Finland through projects IntelliSense (no. 345068), AI4AI (no. 350667), Ferrari (no. 359047), Business Finland and European Commission through project ARCTIC (no. 101139908). The work used experimental facilities of OtaNano Micronova cleanroom and Research Council of Finland Infrastructure "Printed Intelligence Infrastructure" (PII-FIRI, project no. 358618).

Supporting Information

**Record High Polarization at 2V and Imprint-free operation in Superlattice $HfO_2$-$ZrO_2$ by Proper Tuning of Ferro and Antiferroelectricity**

Xinye Li, Sayani Majumdar*

Information Technology and Communication Sciences, Tampere University, 33720 Tampere, Finland

*Email: sayani.majumdar@tuni.fi

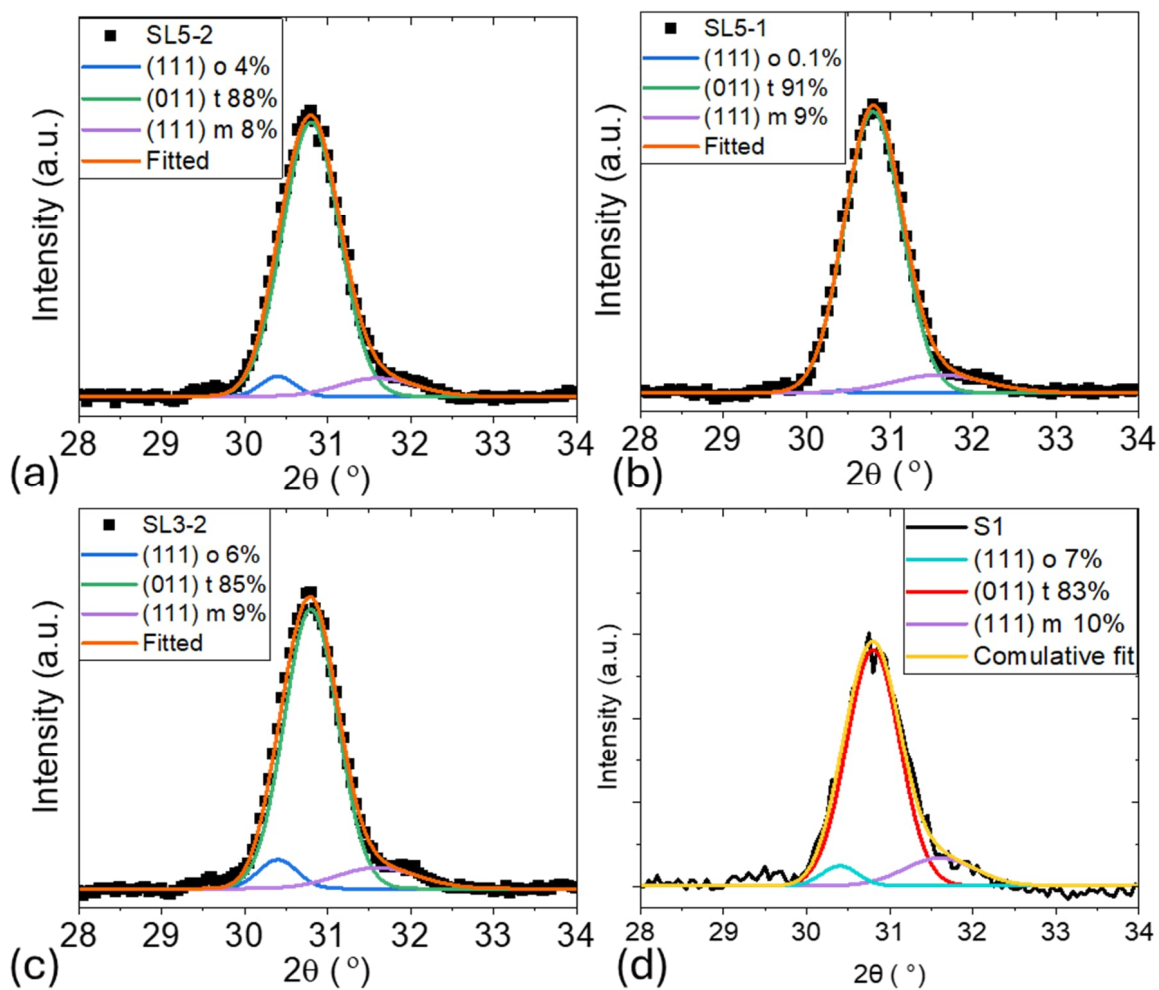

**Figure S1 (a, b, c, d)** GIXRD 2θ scans in range 28° to 34° over the predicted o (111), t (011), m (111) peak position of S1, SL(3,5)-2 and SL5-1. Peak fitting of 2θ scans with different reference diffraction peaks are used to calculate the ratio of the different crystalline phases. Same technique as in [X. Li 2025] is applied here.

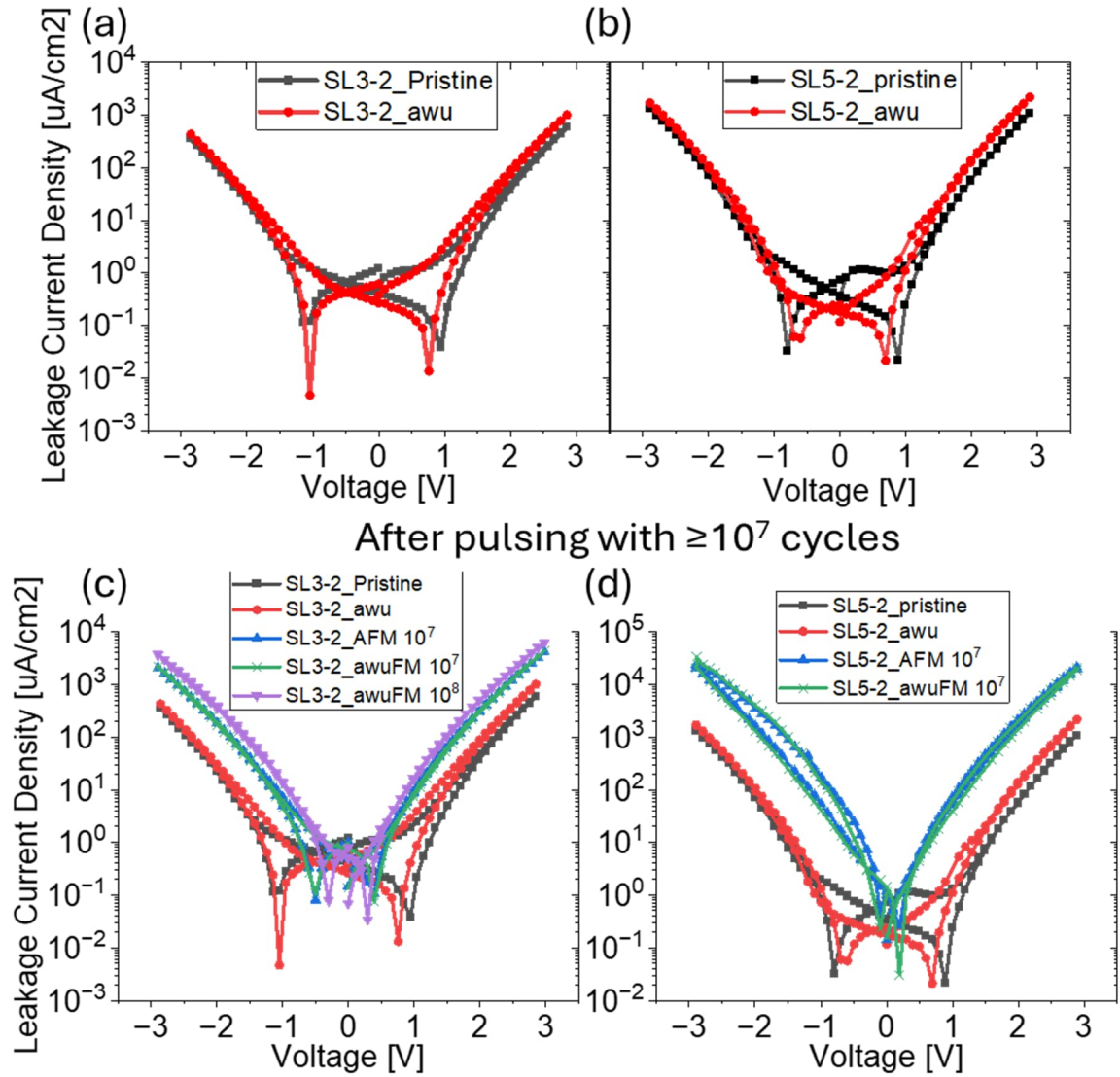

**Figure S2** Static *I-V*s of samples SL3-2 (**a, c**) and SL5-2 (**b, d**), black curves are the pristine state with no electric pulsing applied, blue curves are measured after $10^7$ cycles of endurance test without wake-up pulsing, red curves measured after 1000 cycles of wake-up pulsing, green and purple curves are measured after post wake-up endurance test with specific pulsing cycles.

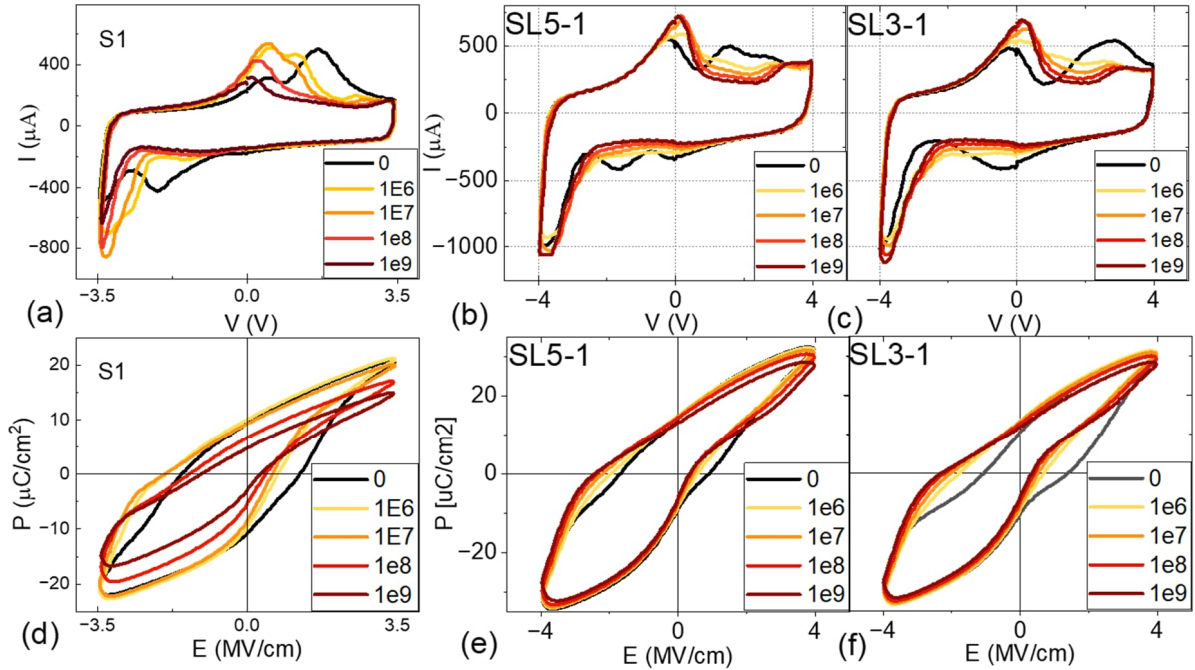

**Figure S3** Evolution of monitoring hysteresis in DHM mode during the endurance test of the PDA-treated samples: *I-V* hysteresis **(a-c)**, *P-E* hysteresis **(d-f)** shape from pristine to fatigued samples, showing a continuous imprint coercive field for both S1 and SL(3, 5)-1 and while visible degradation of switching current only for S1.

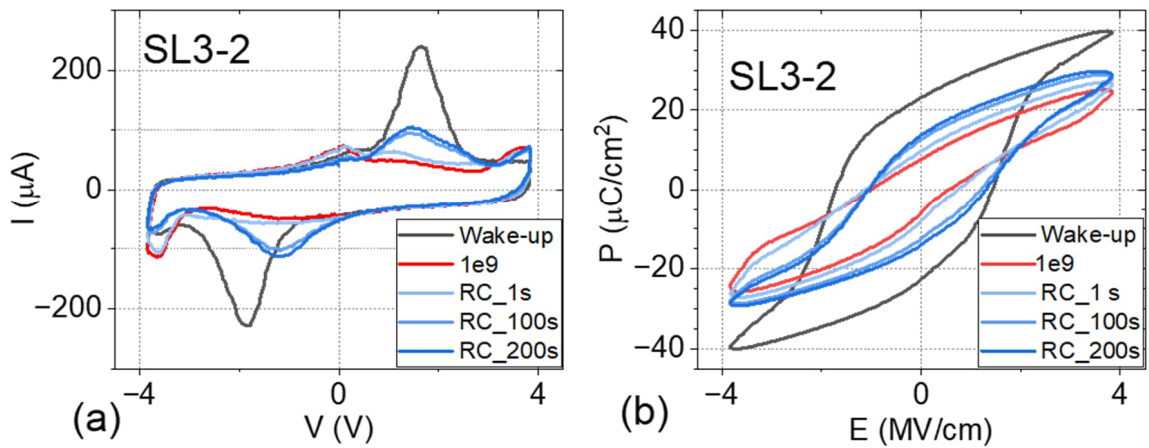

**Figure S4** Woken-up in black curves, fatigued in red curves and recovered hysteresis from 1 s to 200 s in from light to dark blue curves, (a) *I-V,* (b) *P-E* hysteresis of SL3-2. Figure S4 shows an extreme fatigue case of SL3-2 after pulsing for $10^9$ cycles long, it can be recovered to 60% ($2P_r$: 27.85 μC/cm$^2$) of the woken-up state via 200 s of RC pulsing.